# Wavelength tunable spectral compression in a dispersion-increasing fiber

Hsiu-Po Chuang and Chen-Bin Huang*

*Institute of Photonics Technologies, National Tsing Hua University, 101 Sec. 2 Kuang Fu Road, Hsinchu 30013, Taiwan*
*Corresponding author: robin@ee.nthu.edu.tw*

Adiabatic soliton spectral compression in a dispersion-increasing fiber is demonstrated both numerically and experimentally. We show a positively-chirped pulse provides better spectral compression in a dispersion-increasing fiber with large anomalous dispersion ramp. An experimental spectral compression ratio of 15.5 is obtained using 350 fs positively-chirped input pulse centered at 1.5 μm. A 30 nm wavelength tuning ability is experimentally achieved.

Coherent optical sources with high spectral brightness and wide wavelength tuning range are highly desirable in spectroscopic applications, nonlinear microscopy, as well as wavelength-swept optical coherence tomography. The spectral brightness of a wideband optical source can be effectively enhanced through the redistribution of the source energy into a narrower spectral range. A solution is to perform spectral compression. The spectral narrowing effect was first explained for a negatively-chirped optical pulse propagating in optical fibers in the normal dispersive regime [1,2]. It was demonstrated later, chirped optical pulses within standard single-mode fiber [3,4] and photonic crystal fiber [5,6] with anomalous dispersions were also capable of achieving spectral compression. Such spectral narrowing effects have also been observed in normal dispersive gain fiber [7] and photonic crystal fiber [8]. Recently, a comb-profile fiber was used to demonstrate adiabatic soliton spectral compression with a compression ratio up to 26 [9]. However, extreme care in the design and fusion splicing among nineteen concatenations of standard single-mode and dispersion shifted fibers were required.

While successfully observed in the above works, only a few works demonstrated spectral compressions with wide wavelength tuning abilities. In Ref. [9], wavelength tuning was achieved using different input pulse wavelengths rather than during the spectral compression process. In Ref. [10], soliton self-frequency shift was utilized to realize wide (~300 nm) wavelength tunable spectral compression in a highly nonlinear photonic crystal fiber, with spectral compression ratio of 6.5.

Dispersion-decreasing fiber, in which a single fiber segment with gradual dispersion ramp is realized during fiber drawing process, has been widely adopted to achieve adiabatic soliton temporal compression [11] and finds many applications in coherent communications [12] and optical arbitrary waveform generations [13]. Implementing the dispersion-decreasing fiber reversely, a dispersion-increasing fiber (DIF) can be obtained and should enable a straightforward means in accomplishing adiabatic soliton spectral compression. However, such simple approach has not been realized to our best knowledge.

In this Letter, the feasibility of adiabatic soliton spectral compression in a DIF is first assessed numerically. We demonstrate the first (to the best knowledge) experimental spectral compression in a 1-km dispersion flattened DIF. We show a positively-chirped pulse can obtain a larger spectral compression ratio in a DIF with large anomalous dispersion ramp. An experimental spectral compression ratio of 15.5 is obtained. A 30 nm wavelength tuning range of the spectral compressed spectra is experimentally observed. Our experimental results are in excellent accord to calculations.

The dispersion-flattened DIF considered in this paper is 1-km in length. A linear dispersion ramp of 0.6 to 13.5 ps/nm/km (comply with experimental DIF specification) from the fiber input to output is assumed. This permits an ideal spectral compression ratio of 13.5/0.6=22.5, determined by the ratio between the dispersion values at the DIF output to that of the input. The numerical results are obtained by solving the generalized nonlinear Schrödinger equation using split-step Fourier method [14], with 4000 computational steps. A fiber loss coefficient of 0.4 dB/km is used, along with the nonlinear and Raman coefficients of 3.5 $(W \cdot km)^{-1}$ and 3 fs, respectively. The nonlinear coefficient is taken as constant here since the variation in the core size is typically negligible [11]. In our numerical studies, the self-steepening effect is not evident since the pulse duration is much broader than the carrier cycle, and the required peak power for the input pulse is low due to a small input DIF dispersion value.

Figure 1(a) shows the calculated spectral evolutions when a 205 fs fundamental soliton with full-width half-maximum (FWHM) bandwidth of 12.5 nm, centered at 1560 nm, is launched into the DIF. The output spectral FWHM width shown in Fig. 1(a) is 2.8 nm, giving a spectral compression ratio of only 4.5. This is due to the rapid dispersion variation as a function of propagation distance within the 1-km length, hindering ideal adiabatic soliton spectral compression after a propagation of roughly 170 m.

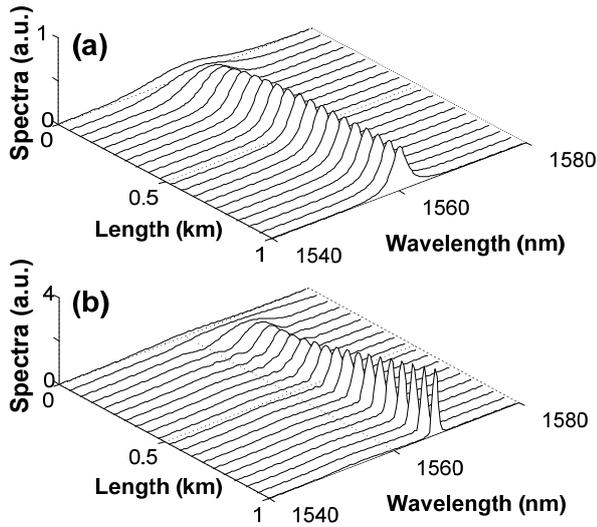

Fig. 1. Calculated evolutions of adiabatic soliton spectral compression for (a) transform-limited and (b) up-chirped pulses.

An enhanced spectral compression ratio can be obtained (given the same dispersion ramp value and fiber length) when positively-chirped input pulse is considered. We hereby provide a brief discussion to the working principle behind such approach: A dispersion ramp ratio of 22.5 within a 1-km fiber length is considered too rapid to sustain adiabatic soliton spectral compression. By providing an initial positive-chirp to the input pulse, the pulse experiences gradual temporal compression as it propagates down the DIF since the anomalous dispersion of the DIF compensates the positive-chirp. After a certain propagation distance, the formation of fundamental soliton is possible when the total anomalous dispersion provided by the DIF completely compensates the initial positive-chirp. At that specific distance, the corresponding dispersion value is increased as compared to the input end, thus permitting stable adiabatic soliton propagation and results in an improved spectral compression ratio.

Our reasoning is corroborated in Fig. 1(b), which shows the calculated spectral evolutions when a 350 fs positively-chirped pulse resulting from the same 12.5 nm optical spectrum is launched into the DIF. The resulting output spectral FWHM width of 1.0 nm gives a much higher spectral compression ratio of 12.5, and the spectral brightness is five-times greater as compared to the transform-limited case. It is found the compression ratio is a function of launched pulse power and can thus be optimized. For chirped input pulses, adiabatic soliton spectral compression is accompanied with evident Raman red-shifts. This property is exploited for wavelength tunable spectral compression to be demonstrated later.

Figure 2(a) shows the schematic experimental setup. The optical source is an Er-doped mode-locked fiber laser (MLFL) with 50 MHz repetition-rate. The direct laser output spectrum with 13 nm FWHM bandwidth is shown in Fig. 2(b). A short-pulse erbium-doped fiber amplifier (EDFA, Pritel LNHPFA-27) is used to provide power tuning ability. The pulses are split into two paths using a 3-dB coupler. A segment of dispersion-compensating fiber (DCF1, providing a quadratic phase coefficient of 8.8e3 fs$^2$) is used so positively-chirped pulses are launched into the DIF. Another segment of dispersion-compensating fiber (DCF2) is used to ensure 205 fs transform-limited pulses are delivered to the reference arm of a home-made intensity cross-correlator. The dispersion flattened DIF (FORC, Moscow) employed in this experiment is 1 km in length, with linear dispersion ramp from 0.6 to 13.5 ps/nm/km. The power launched into the DIF is monitored via a power meter (PM). The spectral and temporal responses after the DIF are measured with an optical spectrum analyzer (OSA, Yokogawa AQ6370C) and the intensity cross-correlator, respectively. Intensity autocorrelation traces of the positively-chirped pulse for the DIF input are shown in Fig. 2(c). The experimental autocorrelation trace (symbols) is plotted against the calculated trace (solid line), giving a 350 fs de-convoluted FWHM duration. In the calculation, we used the experimental MLFL spectrum and the dispersion value provided by the DCF vendor (OFS).

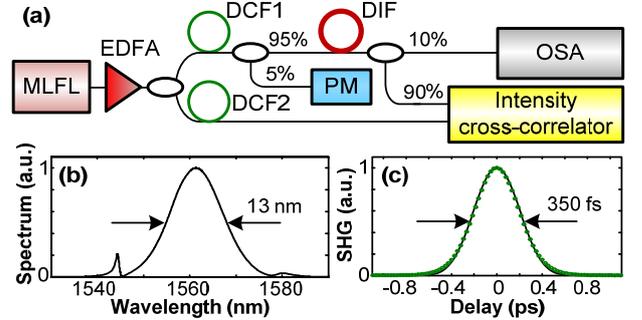

Fig. 2. (Color online) (a) Schematic experimental setup. MLFL: mode-locked fiber laser; DCF: dispersion-compensating fiber; PM: power meter. (b) Initial laser spectrum. (c) Calculated (line) and experimental (symbol) intensity autocorrelation traces of the DIF input pulse.

Figure 3(a) shows our experimental spectrally compressed optical spectrum after the DIF. Here the EDFA gain is adjusted for optimized spectral compression ratio. With an input average power of 0.67 mW, the output spectral FWHM width is 0.84 nm, giving a spectral compression ratio of 15.5. The center wavelength of the compressed spectrum is red-shifted to 1569.5 nm. The experimental parameters are inserted into the simulation and the result is show in Fig. 3(b). The experiment and calculation are in excellent agreement in terms of the FWHM width, center wavelength shift, as well as the spectral pedestal centered at 1557 nm. The experimental (symbol) and calculated (solid line) intensity cross-correlation traces after the DIF are shown in Fig. 3(c), giving a 5.3 ps pulse duration after soliton spectral compression. We attribute the slight asymmetry and the broader shoulder in the experimental cross-correlation trace to the 1-km path difference between the reference and the DIF arms, resulting in non-negligible timing jitter during the correlation measurement.

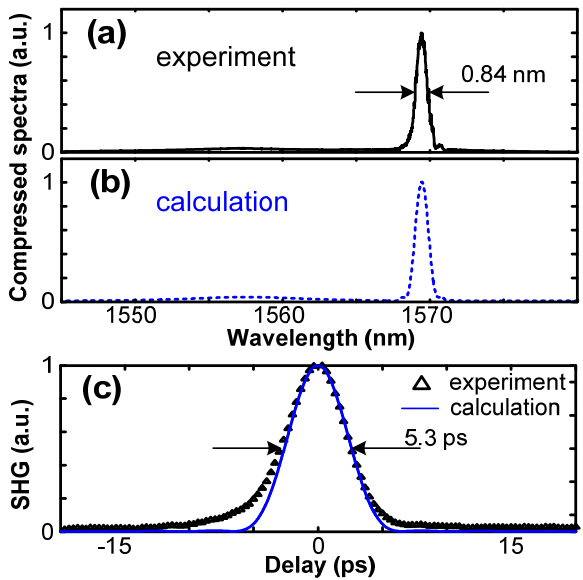

Fig. 3. (Color online) DIF outputs. (a,b) Experimental and calculated compressed spectra. (c) Experimental (symbol) and calculated (solid) intensity cross-correlation traces.

With different input pulse power values, Fig. 4(a) shows our experimental wavelength tunable spectral compression results. By increasing the average power from 0.67 mW to 0.96, 1.23, and 1.32 mW, the spectrally compressed spectra can be correspondingly red-shifted from 1569.5 nm to 1578.7, 1591.7, and 1599.3 nm to achieve a nearly 30 nm tuning range. The experimental results are compared to calculations as shown in Fig. 4(b), and are again found in good accord. We note here, the benefit of wavelength tuning ability is accompanied with sacrificed compression ratio. During the red-shifts, the experimental spectral compression ratios are gradually decreased from 15.5 (0.67 mW) to 7.7 (0.96 mW), 7.2 (1.23 mW), and 5.0 (1.32 mW). However, even the smallest compression ratio of 5 is comparable to a recent spectral compression demonstration in a photonic crystal fiber using amplitude optimized pulses [8].

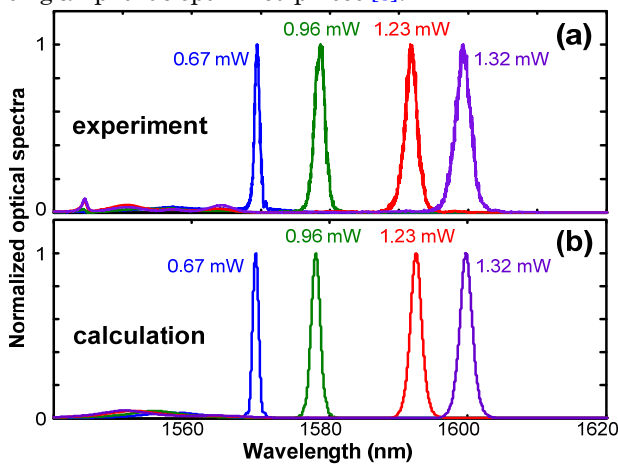

Fig. 4. (Color online) Wavelength tunable spectral compression. (a) Experimental and (b) calculated optical spectra for the average power values labeled.

In summary, adiabatic soliton spectral compression in a dispersion-increasing fiber (DIF) with linear dispersion ramp is demonstrated numerically and experimentally. We show both theoretically and experimentally, a positively-chirped optical pulse can result in a better spectral compression ratio in a DIF with large dispersion ramp as compared to transform-limited input pulse. An experimental spectral compression ratio of 15.5 is achieved using 350 fs positively-chirped pulses centered at 1562 nm using a dispersion-flattened DIF with 0.6 to 13.5 ps/nm/km linear dispersion ramp. A 30 nm wavelength tuning ability is achieved experimentally with different launched pulse power. Our experimental results are in excellent accord to numerical results.

This work was supported by the National Science Council in Taiwan under contract NSC 97-2112-M-007-025-MY3. The authors wish to thank Prof. S-D. Yang for the support on the mode-locked fiber laser.

### References

1. M. Oberthaler and R. A. Hopfel, Appl. Phys. Lett. **63**, 1017 (1993).
2. S. A. Planas, N. L. Pires Mansur, C. H. Brito Cruz, and H. L. Fragnito, Opt. Lett. **18**, 699 (1993).
3. S. Shen, C.-C. Chang, H. P. Sardesai, V. Binjrajka, and A. M. Weiner, J. Lightwave Technol. **17**, 452 (1999).
4. B. R. Washburn, J. A. Buck, and S. E. Ralph, Opt. Lett. **25**, 445 (2000).
5. E. R. Andresen, J. Thøgersen, and S. R. Keiding, Opt. Lett. **30**, 2025 (2005).
6. D. A. Sidorov-Biryukov, A. Fernandez, L. Zhu, A. Pugžlys, E. E. Serebryannikov, A. Baltuška, and A. M. Zheltikov, Opt. Express **16**, 2502 (2008).
7. J. Limpert, T. Gabler, A. Liem, H. Zellmer, and A. Tünnermann, Appl. Phys. B **74**, 191 (2002).
8. E. R. Andresen, J. M. Dudley, D. Oron, C. Finot, and H. Rigneault, Opt. Lett. **36**, 707 (2011).
9. N. Nishizawa, K. Takahashi, Y. Ozeki, and K. Itoh, Opt. Express **18**, 11700 (2010).
10. A. B. Fedotov, A. A. Voronin, I. V. Fedotov, A. A. Ivanov, and A. M. Zheltikov, Opt. Lett. **34**, 662 (2009).
11. K. R. Tamura and M. Nakazawa, Opt. Lett. **26**, 762 (2001).
12. C.-B. Huang, S.-G. Park, D. E. Leaird, and A. M. Weiner, Opt. Express **16**, 2520 (2008).
13. Z. Jiang, C.-B. Huang, D. E. Leaird, and A. M. Weiner, Nat. Photonics **1,** 463 (2007).
14. G. P. Agrawal, *Nonlinear Fiber Optics*, 4th ed. (Academic Press, 2007).